\documentclass[aps,pra,amsmath,amssymb,floatfix,preprint,superscriptaddress,showpacs]{revtex4}
\usepackage{graphicx,multirow}

\newcommand{\co}{{\bf c}}
\newcommand{\bo}{{\bf b}}
\newcommand{\ron}{{\bf r}_1}
\newcommand{\rtw}{{\bf r}_2}

\begin{document}

\title{
Dynamical generalization of a solvable family of two-electron model
atoms with general interparticle repulsion}
\author{T.~A.~Niehaus}
\affiliation{Bremen Center for Computational Materials Science, 
          University of Bremen, 
          D-28359 Bremen, Germany }
\affiliation{German Cancer Research Center,
          Dept. Molecular Biophysics,
          D-69120 Heidelberg, Germany }

\author{S.~Suhai}
\affiliation{German Cancer Research Center,
          Dept. Molecular Biophysics,
          D-69120 Heidelberg, Germany }

\author{N.~H.~March}
\affiliation{Oxford University, Oxford, England}
\date{\today}
\begin{abstract}
Holas, Howard and March [Phys. Lett. A {\bf 310}, 451 (2003)] have
obtained analytic solutions for ground-state properties of a whole
family of two-electron spin-compensated harmonically confined model
atoms whose different members are characterized by a specific
interparticle potential energy u($r_{12}$). Here, we make a start on
the dynamic generalization of the harmonic external potential, the
motivation being the serious criticism levelled recently against the
foundations of time-dependent density-functional theory
(e.g. [J. Schirmer and A. Dreuw, Phys. Rev. A {\bf 75}, 022513
(2007)]). In this context, we derive a simplified expression for the
time-dependent electron density for arbitrary interparticle
interaction, which is fully determined by an one-dimensional non-interacting
Hamiltonian. Moreover, a closed solution for the momentum space density in the
Moshinsky model is obtained. 
\end{abstract}

\maketitle

\section{Background and outline}
A family of two-electron spin-compensated harmonically confined model
atoms has recently been proposed and studied by Holas, Howard and
March \cite{hol03} (referred to below as HHM). Each member of this
family is characterized by a specific interparticle potential energy
$u(|{\bf r}_1-{\bf r}_2|)$. HHM showed that the ground-state spatial wave function $\Psi(\ron,\rtw)$ then separated in centre-of-mass ($\co$) and relative motion ($\bo$) coordinates defined by 
\begin{equation}
\label{vdef}
\co = \frac{\ron + \rtw}{2} \quad,\quad \bo = \ron - \rtw,
\end{equation}
to yield 
\begin{equation}
\label{gscr}
\Psi(\ron,\rtw) =  \psi_{\text{CM}}(\co) \psi_{\text{RM}}(\bo). 
\end{equation}
The centre-of-mass term $\psi_{\text{CM}}$ is determined once for all as a Gaussian function. Naturally the relative motion part $ \psi_{\text{RM}}$ entering Eq.~(\ref{gscr}), involves the interparticle repulsion $u(b)$, but only through an effective one-body potential energy $V_\text{eff}(b)$ given  by (here and in the following  atomic units are used [$\hbar, e, m_e = 1$])
 \begin{equation}
\label{eff}
V_\text{eff}(b) = \frac{1}{4} \omega_0^2 b^2 + u(b),
\end{equation}
which must then be inserted in a one-particle Schr\"{o}dinger equation
to solve for the relative motion wave function  $ \psi_{\text{RM}}$. Analytic
solutions exist for a number of choices of $u(b)$ in Eq.~(\ref{eff}),
including harmonic \cite{moc68}, Coulombic \cite{kaiis}  and inverse
square law \cite{cran,cap05}.

As to the motivation for the dynamic generalization of the HHM static
family of solutions in Ref.~\cite{hol03} above, we cite the recent
criticism of Schirmer and Dreuw \cite{schi} that lies at the foundations of time-dependent density-functional theory (TDDFT); a very popular approximate route for current calculations of electronic excitation energies in atoms and molecules \cite{dre05}. Because of such criticism, it seemed the more important to broaden considerably the class of available exact time-dependent analytical treatments. In that context, we must note here for Coulombic interaction in a two-electron model with harmonic confinement \cite{kas}, the study of D'Amico and Vignale \cite{ami} in which the harmonic confinement is generalized to be time-dependent, the dimensionality of the problem these authors considered being reduced from three in the so-called Hookean atom \cite{kaiis,kas} to two. 

Our objective here is to effect a related generalization of the whole
family of two-electron model atoms considered by HHM to time-dependent
theory. We therefore keep the discussion as general as possible and
turn to specific forms of the interparticle interaction only in
analytically solvable cases. The resulting expressions could serve as
a convenient starting point for a simplified numerical treatment,
which is however outside the scope of this work.

\section{Time propagation of harmonically confined two-electron model
  atoms}
\label{sec1}
We wish to solve the  Schr\"{o}dinger equation
 \begin{equation}
i \frac{\partial}{\partial t} \Psi = \hat{H} \Psi.
\end{equation}  
Here we take the model which is a time-dependent generalization of the study of HHM. Thus , the Hamiltonian operator assumed has the form 
 \begin{equation}
\hat{H} = - \frac{1}{2} \left( \nabla_{r_1}^2 + \nabla_{r_2}^2 \right) + \frac{1}{2} \omega_0^2(t) \left( r_1^2 + r_2^2 \right) + u(|\bf{r}_1 - \bf{r}_2|).
\end{equation} 

The time-dependent external potential involving $  \omega_0^2(t)$ drives the system from its ground state $\Psi_0$ at t=0 to a general time-dependent state $\Psi(\bf{r}_1,\bf{r}_2,t)$ at time t. Following HHM we use centre-of-mass vector $\co$ and relative motion vector $\bo$ in Eq.~(\ref{vdef}), to find 
 \begin{multline}
\label{both}
i \frac{\partial}{\partial t}  \Psi(\bo,\co,t)= \\\left[ -\nabla_{\bo}^2 -  \frac{1}{4} \nabla_{\co}^2 +   \frac{1}{4} \omega_0^2(t) (b^2 + 4c^2) + u(b)\right]  \Psi(\bo,\co,t).
\end{multline} 
With the help of the product ansatz 
\begin{equation}
\label{proda}
 \Psi(\bo,\co,t) =  \psi_{\text{CM}}(\co,t)  \psi_{\text{RM}}(\bo,t)
\end{equation} 
Eq.~(\ref{both}) is without loss of generality readily  separated into relative motion and  centre-of-mass channels according to
\begin{eqnarray}
 &i& \frac{\partial}{\partial t}  \psi_{\text{CM}}(\co,t) = \\\nonumber &&\left[ -  \frac{1}{2m_{\text{cm}}} \nabla_{\co}^2 +  \frac{m_{\text{cm}}}{2}\omega_0^2(t) c^2  \right]  \psi_{\text{CM}}(\co,t) \\
&i& \frac{\partial}{\partial t}  \psi_{\text{RM}}(\bo,t) = \\\nonumber &&\left[ -
  \frac{1}{2m_{\text{rm}}}\nabla_{\bo}^2 +   \frac{m_{\text{rm}}}{2 }\omega_0^2(t) b^2 + u(b) \right]  \psi_{\text{RM}}(\bo,t),
\end{eqnarray} 
with effective masses $m_{\text{cm}}=2$ and $m_{\text{rm}}=1/2$.
It should be noted, that any (time-dependent) separation constant would give rise to phase factors of opposite sign in the RM and CM wave functions, which would then cancel in the formation of the total wave function and is hence neglected. 
\subsection{Solution of the CM problem in two and three dimensions}
Since the centre-of-mass Hamiltonian does not depend on the
interaction potential $u(b)$, the CM system be solved once and for
all. In two dimensions this has already been accomplished by D'Amico and Vignale  \cite{ami} for an arbitrary time dependence of $\omega_0(t)$. Their result for the CM wave function reads
\begin{equation}
\label{2dans}
 \psi^{\text{CM,2D}}(\co,t) =  \sum_{n,m}  c_{nm} \chi_{nm}(c,t)  \Theta_m(\theta),
\end{equation} 
where the fact that the Hamiltonian does not depend on the centre-of-mass angular variable $\theta$ was used to separate the wave function into an angular part characterized by the quantum number $m$
  \begin{equation}
 \Theta_m(\theta) = \frac{1}{\sqrt{2 \pi}} e^{-im\theta},
\end{equation} 
and a radial part
  \begin{equation}
\label{chi2d}
 \chi_{nm}(c,t) = A(t) c^m e^{B(t) c^2} L_n^m[C(t) c^2],
\end{equation} 
which involves the generalized Laguerre polynomials $L_n^m$ and the
purely time-dependent functions \cite{rem}
\begin{eqnarray}
\label{abc1}
A(t) &=&
\sqrt{\frac{n!\,2}{(n+m)!}}\,\,[m_{\text{cm}}\dot{\phi}(t)]^{\frac{m+1}{2}} \nonumber\\&&
   \quad\,\times\quad e^{-
 i (2 n + m + 1)  [\phi(t)-\phi(0)]} 
\\\label{abc2}
B(t) &=&  - \frac{m_{\text{cm}}}{2} \left[\dot{\phi}(t) - i \frac{d\ln  |X(t)|}{dt}\right]\\
\label{abc3}
C(t)  &=&  m_{\text{cm}} \, \dot{\phi}(t) .
\end{eqnarray}
The complex functions $X(t)$
  \begin{equation}
 X(t) = |X(t)|e^{i\phi(t)} \quad\text{with}\quad \dot{\phi}(t) > 0,
\end{equation} 
are solutions to the equation of motion for the classical harmonic oscillator
  \begin{equation}
 \ddot{X} = -\omega_0^2(t) X(t), 
\end{equation} 
which can be solved once an explicit form of $\omega_0^2(t)$ is chosen. Remarkably, the time propagation of the harmonically confined quantum system is fully determined by its classical analogue.

As a new result, we compute in the following the CM wave function  for
the general case of three dimensions. The derivation turns out to be
straightforward and parallels the 2D case with minor modifications. To
start with, we write the state $\psi^{\text{CM,3D}}_{nlm}$ with main quantum number $n$, angular momentum $l$ and magnetic quantum number $m$ as a product of a radial part  $\tilde{\chi}_{nl}$ and spherical harmonics $Y_{lm}$:
\begin{equation}
\label{ansn}
 \psi_{nlm}^{\text{CM,3D}}(\co,t) =  \tilde{\chi}_{nl}(c,t)  Y_{lm}(\theta,\phi).
\end{equation} 

The more general case in which the system is not in one of its
eigenstates at $t=0$ can be handled easily according to 
Eq.~(\ref{2dans}), since the expansion coefficients do not depend on
time. Notwithstanding, the ansatz Eq.~(\ref{ansn}) fully allows for
excitations to eigenstates with different main quantum number due to the
time dependent potential realized by $\omega_0(t)$.
 
Proceeding,  the radial equation of motion takes the following
form:
\begin{widetext}
 \begin{equation}
  \label{3drad}
 i \frac{\partial}{\partial t}  \tilde{\chi}_{nl}(c,t) = \left[  -
   \frac{1}{2 m_{\text{cm}}} \frac{\partial^2}{\partial c^2} -
   \frac{1}{m_{\text{cm}} c}
   \frac{\partial}{\partial c} + \frac{l(l+1)}{2 m_{\text{cm}} c^2} + \frac{1}{2}m_{\text{cm}}\omega_0^2(t) c^2  \right] \tilde{\chi}_{nl}(c,t), 
\end{equation} 
\end{widetext}

which we solve using a time-dependent generalization of the well known result for the
isotropic harmonic oscillator in the ground state
\begin{equation}
  \label{3dans}
   \tilde{\chi}_{nl}(c,t) = \tilde{A}(t) c^l e^{\tilde{B}(t) c^2}
     L^{l+1/2}_{\frac{1}{2}(n-l)} [ \tilde{C}(t) c^2]. 
\end{equation}

Inserting Eq.~(\ref{3dans}) into Eq.~(\ref{3drad}) and taking
advantage of the defining differential equation of the associated
Laguerre polynomials, we obtain after some algebra the following expressions for the
functions $\tilde{A}, \tilde{B}$ and $\tilde{C}$:
\begin{align}
  \label{3dabc1}
  &i \dot{\tilde{A}} + \frac{(2l+3)}{m_{\text{cm}}} \tilde{A} \tilde{B} -
    \frac{(n-l)}{m_{\text{cm}}}\tilde{A} \tilde{C} = 0\\
  &i \dot{\tilde{B}} + \frac{2}{m_{\text{cm}}}\tilde{B}^2 - \frac{1}{2} m_{\text{cm}}\omega_0^2(t)  = 0\\\label{3dabc2}
 &i \dot{\tilde{C}} + \frac{4}{m_{\text{cm}}} \tilde{B} \tilde{C} +
 \frac{2}{m_{\text{cm}}} \tilde{C}^2  = 0.
\end{align}
Equations (\ref{3dabc1}) to (\ref{3dabc2}) are (besides different
prefactors) completely equivalent to equations A4 to A6 in the
mentioned work of D'Amico and Vignale, which allows us to write down
the radial solution in three dimensions immediately:
\begin{multline}
  \label{3dres}
   \tilde{\chi}_{nl}(c,t) = \sqrt{\frac{2^{n+l+2} m_{\text{cm}}^{l+\frac{3}{2}}
       [\frac{1}{2}(n-l)]! [\frac{1}{2} (n+l)]!
        }{\sqrt{\pi} (n+l+1)!}} 
    \quad \times \\ [\dot{\phi}(t)]^{\frac{2l+3}{4}} \exp\left[-\frac{1}{2} m_{\text{cm}} (\dot{\phi}(t) - i \frac{d\ln  |X(t)|}{dt}) c^2\right] \quad\times \\
   e^{-
 i (n +\frac{3}{2} )  [\phi(t)-\phi(0)]}  c^l    L^{l+1/2}_{\frac{1}{2}(n-l)} [m_{\text{cm}} \dot{\phi}(t)  c^2].
  \end{multline}
   
\subsection{Time dependent electron density for general interparticle interaction}
Having solved the problem for the CM system we now turn to
an evaluation of the time-dependent electron density $n({\bf r},t)$
for the special but important case of a system that is in its ground
state at $t=0$. The square modulus of the CM wave function then reduces
to a simple Gaussian function
\begin{equation}
  \label{sm}
  |\psi_{000}^{\text{CM,3D}}(\co,t)|^2 = \frac{1}{a^3_{\text{CM}}(t) \pi^{3/2}} \exp\left(-\frac{c^2}{a^2_{\text{CM}}(t)}\right),
\end{equation}
where the time-dependence is fully governed by the characteristic
length scale $a_{\text{CM}}(t)$ of the oscillator:
  \begin{equation}
    \label{acm}
    a_{\text{CM}}(t) = \frac{1}{m_{\text{cm}} \dot{\phi}(t)}.
  \end{equation}
Since the potential $u(b)$ depends on the interparticle distance only,
also the relative motion wave function $\psi^{RM,3D}$ can once again be split into
angular and radial parts according to 
\begin{equation}
  \label{rmwf}
  \psi_{nlm}^{\text{RM,3D}}(\bo,t) =  \tilde{\zeta}_{nl}(b,t)
  Y_{lm}(\theta,\phi), 
\end{equation}
where $\tilde{\zeta}_{nl}(b,t)$ needs in general to be determined by
time-propagation in one spatial dimension.

For the mentioned special case the electron density  is
given by
\begin{multline}
  \label{dns1}
  n({\bf r_1},t) =\\ 2 \int \left| \psi_{000}^{\text{CM,3D}}(\frac{1}{2}({\bf r_1}+{\bf r_2}),t)
    \psi_{000}^{\text{RM,3D}}({\bf r_1}-{\bf r_2},t)\right|^2 d{\bf r}_2,
\end{multline}
which can be considerably simplified using Eq.~(\ref{sm}) and
(\ref{rmwf}). After evaluation of the angular integrations and
substitution of ${\bf y} = ({\bf r_1}-{\bf r_2})/a_{\text{CM}}$ for ${\bf r_2}$,
we arrive at
\begin{eqnarray}
  \label{dnsf}\nonumber
  n({\bf r_1},t) &=& \frac{8}{\sqrt{\pi}}
  \exp(-\frac{r_1^2}{a^2_{\text{CM}}(t)}) \\\nonumber &&\times
\int_0^\infty dy\,\, y^2  \exp(-\frac{y^2}{4}) 
 \left|\psi_{000}^{\text{RM,3D}}(a_{\text{CM}}(t) y,t)\right|^2 
 \\ && \times \frac{\sinh(r_1
   y/a_{\text{CM}}(t))}{(r_1 y/a_{\text{CM}}(t))} ,
\end{eqnarray}
which constitutes a non-trivial generalization of the HHM result for
the static ground state density given in Eq.~(14) of that publication.    

\subsection{Time-dependent atomic scattering factor}
A quantity which is easily accessible by experiment is the atomic
scattering factor given by the Fourier transform of the atomic electron
density
\begin{equation}
\label{scat}
  f({\bf k},t) = \int n({\bf r},t) e^{i{\bf k} {\bf r}} d{\bf r}.
\end{equation}

Inserting Eq.~(\ref{proda}) into Eq.~(\ref{scat}) and taking advantage
of the fact that the Jacobian for the transformation $\{\ron,\rtw\}
\rightarrow \{\bo,\co\}$ is unity, we obtain 
 \begin{eqnarray}
\label{scat2}
  f_{\text{tot}}({\bf k},t) &=& 2\,\left[\int |\psi_{\text{CM}}(\co,t)|^2
    e^{i{\bf k}\co}  d\co \right]\nonumber\\&&\times \left[\int|\psi_{\text{RM}}(\bo,t)|^2
  e^{i\frac{{\bf k} \bo}{2}}\,d\bo\right]\nonumber\\
&:=& 2 \,f_{\text{\text{CM}}}({\bf k},t)\, f_{\text{\text{RM}}}({\bf k}/2,t),
\end{eqnarray}
which shows that the total two-electron scattering factor
$f_{\text{tot}}$ decouples into a product of one-particle scattering
factors of the centre-of-mass
and relative motion systems. 

While the term $f_{\text{RM}}$ remains until the relative motion
Schr\"{o}dinger equation is solved for a specific $\omega_0(t)$ and
interparticle potential $u(b)$, the other piece can be evaluated
based on the results of the last section. In three dimensions a closed
solution is however only possible for the case of vanishing angular
momentum. We therefore step back to the two dimensional problem studied
by D'Amico and Vignale. For sake of simplicity we treat only the case
for which the system is in an arbitrary eigenstate at $t=0$. The more
general superposition of Eq.~(\ref{2dans}) can be solved along the
lines of the following derivation and poses no additional problems.

Using Eq.~(\ref{chi2d}) the
centre-of-mass structure factor reads
\begin{equation}
  \label{fcm1}
 f_{\text{\text{CM}}}({\bf k},t)   \equiv f_{\text{\text{CM}}}(k,t) = \int_0^\infty J_0(k c)
   |\chi_{nm}(c,t)|^2 c dc,
\end{equation}
where the angular integration led to the Bessel function of the first
kind $J_0$. After insertion of  Eq.~(\ref{abc1}) to (\ref{abc3}) and
polynomial expansion of the Laguerre functions according to
\begin{equation}
  L^m_n(x) = \sum_{s=0}^n {n+m \choose n-s} \frac{(-x)^s}{s!},
\end{equation}
we arrive at
\begin{eqnarray}
  \label{fcm2}\nonumber
  f_{\text{\text{CM}}}(k,t) &=&  \frac{ n!\, 2
  }{(n+m)!}\,\,[m_{\text{cm}}\dot{\phi}(t)]^{m+1}\\\nonumber&\times& \sum_{s=0}^n \sum_{t=0}^n {n+m
    \choose n-s} {n+m
    \choose n-t}  \frac{(-m_{\text{cm}}\dot{\phi})^{s+t}}{s! t!}\\&\times& \int_0^\infty J_0(k c) e^{-m_{\text{cm}}\dot{\phi}(t) c^2}
   c^{2(m+s+t)+1} dc.
\end{eqnarray}
The solution of the remaining integral is known \cite{abra} and
involves the confluent hypergeometric functions of the Kummer type
$M$:
\begin{multline}
 \int_0^\infty  e^{-a^2 t^2} t^{\mu-1} J_\nu(bt)\, dt = \\
 \frac{\Gamma(\frac{\mu+\nu}{2}) \left( \frac{b}{2a} \right)^\nu}{2 a^\mu \Gamma(\nu+1)} M(\frac{\mu+\nu}{2}
 ,\nu+1,-\frac{b^2}{4 a^2}) \\\forall\quad \Re(\mu+\nu)>0;\quad \Re(a^2)>0. 
 \end{multline}
The CM structure factor then takes the final form
\begin{multline}
  \label{finCM}
  f_{\text{\text{CM}}}(k,m_{\text{cm}}\dot{\phi}(t)) = \frac{ n!
  }{(n+m)!} \sum_{s=0}^n \sum_{t=0}^n {n+m
    \choose n-s} {n+m
    \choose n-t} \\ \times \frac{\Gamma(m+s+t+1)}{s!\,t!}
  M(m+s+t+1,1,-\frac{k^2}{4 m_{\text{cm}}\dot{\phi}}),
\end{multline}
which constitutes one of the main results of this work. As indicated, the 
time-dependence of Eq.~(\ref{finCM}) is solely determined by the phase
derivative $\dot{\phi}$, which equals the frequency  $\omega_0$ of the confining
harmonic potential in the static limit. For a system
initially in its ground state, $f_{\text{\text{CM}}}$ reduces to a simple
Gaussian which correctly tends, respectively,  towards the number of electrons
(one in this case) as $k$ approaches zero, and zero as $k$ grows to
infinity. 

Having obtained a closed solution for the centre-of-mass structure
factor we now briefly discuss the relative motion term. In general it will be
necessary to evaluate this part by numerical methods, which is beyond
the scope of the present article. The  so-called Moshinsky
atom \cite{moc68} characterized by the harmonic interparticle potential $u(b)=
-\frac{1}{2} {\cal K} b^2$ provides a useful exception and can be treated analytically. To this end we
note that in this case the relative motion Hamiltonian is equivalent to a
centre-of-mass one with effective mass $\tilde{m}_{\text{cm}} = 1/2$ and
force constant $ \omega_0^2(t) - {\cal K}/ \tilde{m}_{\text{cm}} $. Solving
\begin{equation}
  \label{goo}
\tilde{X}(t) = |\tilde{X}(t)|e^{i\tilde{\phi}(t)}
\quad\text{with}\quad  \ddot{\tilde{X}} = - \left(\omega_0^2(t) -
  \frac{{\cal K}}{\tilde{m}_{\text{cm}}}\right) \,
   \tilde{X},
\end{equation}
allows one to obtain the {\em total} structure factor of the Mochinsky
atom from
\begin{equation}
  \label{moc}
 f_{\text{tot}}^{\cal K}(k,t) = 2
 f_{\text{\text{CM}}}(k,m_{\text{cm}}\dot{\phi}(t))\, f_{\text{\text{CM}}}(k/2,\tilde{m}_{\text{cm}}\dot{\tilde{\phi}}(t)).
\end{equation}

\section{Relation to  TDDFT}

As mentioned in the Introduction, this work was motivated by the
serious criticism of Schirmer and Dreuw \cite{schi} of the 1984
theorems of Runge and Gross \cite{run}. In Ref.~\cite{schi} the proof
in \cite{run} was not only challenged but seemingly refuted. To be
specific, the authors claimed that the variational derivation of the
time-dependent Kohn-Sham equations in Ref.\ \cite{run} is invalid due to
an ill-defined action functional presented there. A nonvariational
formulation would also run into problems, since in this case the
Kohn-Sham system would allow one to reproduce but not to predict the
exact electron density.    

The results of this work can in principle be used to investigate this
issue in an actual numerical calculation. For the two-electron
spin-compensated system at hand, it is possible to construct the exact
exchange-correlation potential from the known electron density in real
(Eq.~\ref{dnsf}) or momentum space (Eq.~\ref{moc}). This is in line
with van Leeuwen's proof \cite{van} of mapping from densities to potentials
in TDDFT with general two-particle interactions. Such a construction
was for example already performed by Lein and K\"{u}mmel \cite{lei}
or D'Amico and Vignale \cite{ami}. Using this potential in the numerical
solution of the time-dependent Kohn-Sham orbitals would then open the
opportunity to compare the propagated and exact electron density.

\section{Summary and possible future directions}

As set out in the Introduction, HHM \cite{hol03} proposed and solved
the problem of a two-electron spin-compensated family of harmonically
confined atoms with a general interparticle  repulsive potential
$u(r_{12})$. 

Using the available time-dependent study of D'Amico and Vignale
\cite{ami}, but now in two dimensions, we display here the
time-dependent density  $n({\bf r},t)$ for a system initially in the
ground state in three dimensions. While the general result is somewhat
formal, for the so-called Moshinsky atom characterized by the choice
$u(b)= -\frac{1}{2} {\cal K} b^2 $ with ${\cal K}$ measuring
the strength of the interparticle repulsion, we derive the momentum
space electron density $f^{\cal K}(k,t)$ functionally from the D'Amico
and Vignale form for the independent two-electron case with ${\cal K}$
set equal to zero. 

For the future, progress may come by taking a specific choice of the
time-dependence $\omega_0^2(t)$ of the harmonic confinement
potential. Singling out a specific Fourier component through the
choice $\exp{i\omega t} + \exp-{i\omega t}$ is restrictive to periodic
time-dependence but may allow, in the future a more elegant discussion
of the  time-dependent particle density
$n({\bf r},t)$ at the heart  of all current theories of TDDFT. But, of
course, more important would be to pass from a model 'He-like' family
to a four-electron interacting 'Be-like' system, where an elegant
Dirac idempotent density matrix already exists in the independent
particle (Hartree-Fock or Kohn-Sham) limit (see, for example
Refs.~\cite{daw,hmt}). 
\section{Acknowledgement}  
 
NHM wishes to thank Professor A.~Holas and I.A.~Howard for invaluable
discussions on the general area embraced by this study. Professor
J.~Schirmer is also thanked by NHM for allowing him early access to
the material in Ref.~\cite{schi}. TAN and NHM acknowledge a generous
scholarship from the German Cancer Research Center, Heidelberg, which
made their contribution to this work possible. 

\end{document}